\documentclass[12pt]{osajnl}

\setboolean{shortarticle}{false} % true = letter, false = research article

\title{Shaping the focal field of radially/azimuthally polarized phase vortex with Zernike polynomials }

\author[*]{Lei Wei}
\author{H. Paul Urbach}

\affil{Department of Imaging Physics, Delft University of Technology, Lorentzweg 1, 2628CJ,  Delft, The Netherlands}

\affil[*]{Corresponding author: l.wei-11@tudelft.nl}

\dates{Compiled \today}

\begin{abstract}
 The focal field properties of radially/azimuthally polarized Zernike polynomials are studied. A method to design the pupil field in order to shape the focal field of radially or azimuthally polarized phase vortex is introduced.  With this method, we are able to obtain a pupil field to achieve a longitudinally polarized hollow spot with a depth of focus up to $12\lambda$ and $0.14\lambda$ lateral resolution for a optical system with numerical aperture 0.99;  A pupil field to generate 8  focal spots along the optical axis is also obtained with this method. 
\end{abstract}

\begin{document}

\maketitle
\ifthenelse{\boolean{shortarticle}}{\abscontent}{}

\section{Introduction}

Optical focal field shaping by engineering the polarization, amplitude and phase on the exit pupil of an optical system\cite{TGBrown00, QWZhan02, QWZhan09}, especially with the help of spatial light modulators \cite{WHan13, JPDing15, CSGuo14}, has attracted lots of attentions in recent years. And this can find applications in many areas. For instance, a beam with phase vortice or azimuthal polarization on the exit pupil can generate hollow focus\cite{TVSPillai2012, CDenz2012}, with null intensity in the center, which can be used to trap absorbing particles, cold atoms\cite{MSZhan2010} and also in high resolution STED microscopy\cite{SWHell2006} . The ability to control the local electric/magnetic field distribution in the focal region makes it possible to determine the orientation of a single optical emitter\cite{SWHell2001} or excite certain resonances of a quantum emitter\cite{LNovotny2015} which is not possible otherwise. There are various methods to shape the focal field of a radially/azimuthally polarized pupil, including designing binary phase masks\cite{HFWang08} or reversing the electrical dipole array radiation\cite{JMWang2010, JMWang2013, YZYu2015}.  Tight focusing behavior of polarized phase vortex beams \cite{LEHelseth2004, CJRSheppard2014} has been investigated to achieve sharp resolution \cite{  XLiu2010, YKozawa2014}, and to demonstrate spin-to-orbital angular momentum conversion\cite{YQZhao07}, et al..  In a previous publication\cite{LWei2014} , we have demonstrated  a method to generate elongated focal spot using Zernike polynomials.  Zernike polynomials forms a complete and orthogonal set of polynomials on a unit circle, which form an ideal set to study the pupil engineering of normally circularly apertured optical imaging system. By precalculating the focal fields of Zernike polynomials, we can optimise the focal field by only optimising the Zernike coefficients.   However, in  paper\cite{LWei2014}, scalar diffraction integral is considered, thus neither polarization nor phase vortices are investigated.   In this paper, we first study the focal field properties of radially/azimuthally polarized Zernike polynomials. Based on this, we applied the pupil engineering method using complex Zernike polynomials to shape the focal field of a phase vortex with radial and azimuthal polarizations to generate an longitudinally polarized axially uniform hollow focus, a transversally polarized elongated spot, and circularly polarized multiple focal spot.\\

\section{Focal field of radial and azimuthal polarized Zernike polynomials}
\begin{figure}[htb]
\centering
{\includegraphics[width=7.5cm]{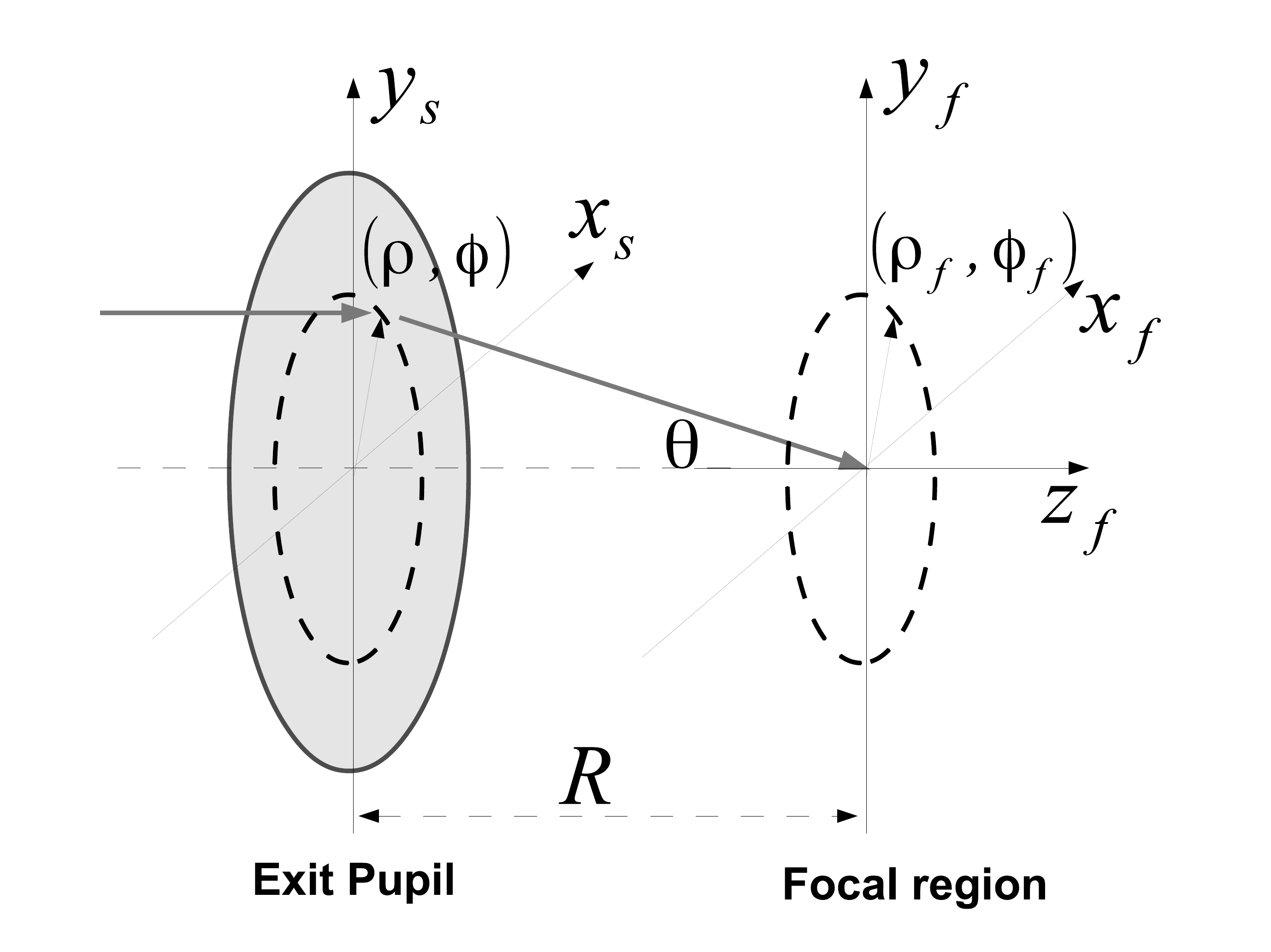}}
\caption{The Debye diffraction model. Light is focused from the exit pupil plane $(x_{s}, y_{s} )$ to the focal region. The focal coordinates is $(x_{\mathrm{f}}, y_{\mathrm{f}}, z_{\mathrm{f}})$ with the geometrical focal point as its origin.}
\label{fig:vectorialDebye}
\end{figure}
The complex Zernike polynomials $Z_n^m(\rho,\phi)$ form a complete set of orthonormal polynomials defined on a unit disc:
\begin{equation}
  Z_n^{m}(\rho,\phi)= R_n^{|m|}(\rho)\mathrm{e}^{\mathrm{i}m\phi},\,
\end{equation}
\begin{equation}
R_{n}^{m}(\rho)=\sum_{s=0}^{p}\frac{(-1)^{s}(n-s)!}{s!(q-s)!(p-s)!}\rho^{n-2s},
\end{equation}
where $ p=\frac{1}{2}(n-|m|),\,q=\frac{1}{2}(n+|m|)$, $n-|m|\ge0$ and even, $(\rho,\phi)$ are the normalised polar coordinates on the exit pupil plane. As we can see from its definition, $Z_n^{m}$ is a phase vortex with the topological charge $m$.  \\
The polarized pupil function can be decomposed into Zernike polynomials:
\begin{equation}\label{eq:zernike}
 \vec{E_s}(\rho,\phi)=\sum_{n,m}\hat{e_s}(\rho,\phi)\beta_n^m Z_n^{m}(\rho,\phi),\,
\end{equation}
for radial polarization there holds that $\hat{e_s}(\rho,\phi)=\hat{e}_{\rho}(\rho,\phi)=(\cos\phi, \sin\phi)$, while for azimuthal polarization $\hat{e_s}(\rho,\phi)=\hat{e}_{\phi}(\rho,\phi)=(-\sin\phi, \cos\phi)$.\\
Zernike polynomials have been very useful to analyze the aberrations of an optical system and have been used as a basic set in the Extended Nijboer Zernike theory, in which a semi-analytical solution of the focal field of each Zernike polynomial on the exit pupil is derived\cite{JJMBraat08, SHaver2013}. \\

In this paper we  compute  the through-focus fields of each radially or azimuthally polarized Zernike polynomial by numerically computing the vectorial Debye diffraction integral\cite{EWolf1958} with Fast Fourier Transform method\cite{TLasser2006}:
\begin{align}\label{eq:debye}
\vec{E}^{nm}_{\mathrm{f}}(x_{\mathrm{f}},y_{\mathrm{f}},z_{\mathrm{f}})&=-\frac{\mathrm{i}R}{2\pi}\int\!\!\!\int_{\Omega}\frac{\sqrt{\cos\theta}\hat{e}_{\mathrm{f}}Z_n^{m}(\rho,\phi)\mathrm{e}^{\mathrm{i}k_{z}z}}{k_{z}}\\ \nonumber
&\times\exp[-\mathrm{i}(k_{x}x_{\mathrm{f}}+k_{y}y_{\mathrm{f}})]\,\mathrm{d}k_{x}\,\mathrm{d}k_{y}
\end{align}
where $R$ is the focal length, the geometrical cone $\Omega$ is defined by $\{(k_x,k_y): k_x^2+k_y^2\le k_0^2 \mathrm{NA}^2\}$,  $k_z=\sqrt{k_0^2n^2-k_x^2-k_y^2}$, $k_0=2\pi/\lambda$, and$(x_{\mathrm{f}},y_{\mathrm{f}},z_{\mathrm{f}})$ is a point in the focal region, with geometrical focal point as origin. For radial polarization, $\hat{e}_{\mathrm{f}}=(\cos\phi\cos\theta, \sin\phi\cos\theta,\sin\theta)$, for azimuthal polarization, $\hat{e}_{\mathrm{f}}=(-\sin\phi, \cos\phi,0)$.\\

Since the vectorial Debye diffraction integral is a linear operator of pupil field $\vec{E_s}$, the focal fields can be expressed by summation of the focal fields of each complex Zernike mode $\vec{E}^{nm}_{\mathrm{f}}$ with the same set of Zernike coefficients $\beta_n^m$ which as occur in the expression of the pupil field:
\begin{equation}
\vec{E}_{\mathrm{f}}(x_{\mathrm{f}},y_{\mathrm{f}},z_{\mathrm{f}})=\sum_{n,m}\beta_n^m \vec{E}^{nm}_{\mathrm{f}}(x_{\mathrm{f}},y_{\mathrm{f}},z_{\mathrm{f}}),
\end{equation}
Once a set of coefficients $\beta_n^m$ for $\vec{E}^{nm}_{\mathrm{f}}$ is found to give a desired focal distribution, the pupil field that gives this focal distribution can be obtained from Eq.(\ref{eq:zernike}).\\
In order to gain more insight into the focal fields of radially and azimuthally polarized Zernike polynomial, we rewrite Eq.(\ref{eq:debye}) using polar coordinates:
\begin{align}
\vec{E}^{nm}_{\mathrm{f}}(\rho_{\mathrm{f}},\phi_{\mathrm{f}},z_{\mathrm{f}})&=-\frac{\mathrm{i}R s_0^2}{\lambda}\int_0^1(1-\rho^2 s_0^2)^{-1/4}\mathrm{e}^{-\mathrm{i}k_0z_{\mathrm{f}}\sqrt{1-\rho^2s_0^2}}\\  \nonumber
&\times R_n^{|m|}(\rho)\rho\mathrm{d}\rho\int_0^{2\pi}\hat{e}_{\mathrm{f}}\mathrm{e}^{\mathrm{i}m\phi}\mathrm{e}^{\mathrm{i}2\pi\rho\rho_{\mathrm{f}}\cos(\phi_{\mathrm{f}}-\phi)}\mathrm{d}\phi,
\end{align}
where, $(\rho_{\mathrm{f}},\phi_{\mathrm{f}},z_{\mathrm{f}})$ are the cylindrical coordinates of a point in the focal region, and $s_0$ is the numerical aperture.\\
Integrals over azimuthal angle $\phi$ can be computes analytically. The cartesian components of the focal electric field of the radially polarized pupil field are given by:
\begin{align}
E_{\mathrm{f},x}^{nm}(\rho_{\mathrm{f}},\phi_{\mathrm{f}},z_{\mathrm{f}})&=-\frac{\mathrm{i}\pi R s_0^2}{\lambda}(-\mathrm{i})^{m+1} \mathrm{e}^{\mathrm{i}m\phi_{\mathrm{f}}}\int_0^1 \rho\mathrm{d}\rho\\ \nonumber
&\times (1-\rho^2s_0^2)^{1/4}\mathrm{e}^{-\mathrm{i}k_0z_{\mathrm{f}}\sqrt{1-\rho^2s_0^2}} R_n^{|m|}(\rho) \\ \nonumber
& \times\left[\mathrm{e}^{\mathrm{i}\phi_{\mathrm{f}}}J_{m+1}(2\pi\rho\rho_{\mathrm{f}})-\mathrm{e}^{-\mathrm{i}\phi_{\mathrm{f}}}J_{m-1}(2\pi\rho\rho_{\mathrm{f}})\right],
\end{align}
\begin{align}
E_{\mathrm{f},y}^{nm}(\rho_{\mathrm{f}},\phi_{\mathrm{f}},z_{\mathrm{f}})&=-\frac{\mathrm{i}\pi R s_0^2}{\lambda}(-\mathrm{i})^{m+2} \mathrm{e}^{\mathrm{i}m\phi_{\mathrm{f}}}\int_0^1 \rho\mathrm{d}\rho \\ \nonumber
& \times(1-\rho^2s_0^2)^{1/4}\mathrm{e}^{-\mathrm{i}k_0z_{\mathrm{f}}\sqrt{1-\rho^2s_0^2}} R_n^{|m|}(\rho)\\ \nonumber
&\times\left[\mathrm{e}^{\mathrm{i}\phi_{\mathrm{f}}}J_{m+1}(2\pi\rho\rho_{\mathrm{f}})+\mathrm{e}^{-\mathrm{i}\phi_{\mathrm{f}}}J_{m-1}(2\pi\rho\rho_{\mathrm{f}})\right],
\end{align}

\begin{align}
E_{\mathrm{f},z}^{nm}(\rho_{\mathrm{f}},\phi_{\mathrm{f}},z_{\mathrm{f}})&=-\frac{\mathrm{i}2\pi R s_0^2}{\lambda}(-\mathrm{i})^{m} \mathrm{e}^{\mathrm{i}m\phi_{\mathrm{f}}}\int_0^1\frac{s_0\rho}{(1-\rho^2s_0^2)^{1/4}}\\ \nonumber
&\times\mathrm{e}^{-\mathrm{i}k_0z_{\mathrm{f}}\sqrt{1-\rho^2s_0^2}} R_n^{|m|}(\rho)J_{m}(2\pi\rho\rho_{\mathrm{f}})\rho\mathrm{d}\rho.
\end{align}
Similarly, the Cartesian coordinates of the azimuthal components of the focal field of an azimuthal polarized pupil field is:
\begin{align}\label{eq:azimex}
E_{\mathrm{f},x}^{nm}(\rho_{\mathrm{f}},\phi_{\mathrm{f}},z_{\mathrm{f}})&=-\frac{\mathrm{i}\pi R s_0^2}{\lambda}(-\mathrm{i})^{m} \mathrm{e}^{\mathrm{i}m\phi_{\mathrm{f}}}\int_0^1\rho\mathrm{d}\rho\\ \nonumber
& \times(1-\rho^2s_0^2)^{-1/4}\mathrm{e}^{-\mathrm{i}k_0z_{\mathrm{f}}\sqrt{1-\rho^2s_0^2}} R_n^{|m|}(\rho)\\ \nonumber
& \times\left[\mathrm{e}^{\mathrm{i}\phi_{\mathrm{f}}}J_{m+1}(2\pi\rho\rho_{\mathrm{f}})+\mathrm{e}^{-\mathrm{i}\phi_{\mathrm{f}}}J_{m-1}(2\pi\rho\rho_{\mathrm{f}})\right],
\end{align}
\begin{align}\label{eq:azimey}
E_{\mathrm{f},y}^{nm}(\rho_{\mathrm{f}},\phi_{\mathrm{f}},z_{\mathrm{f}})&=-\frac{\mathrm{i}\pi R s_0^2}{\lambda}(-\mathrm{i})^{m+1} \mathrm{e}^{\mathrm{i}m\phi_{\mathrm{f}}}\int_0^1\rho\mathrm{d}\rho\\ \nonumber
&\times(1-\rho^2s_0^2)^{-1/4}\mathrm{e}^{-\mathrm{i}k_0z_{\mathrm{f}}\sqrt{1-\rho^2s_0^2}} R_n^{|m|}(\rho)\\ \nonumber
& \times\left[\mathrm{e}^{\mathrm{i}\phi_{\mathrm{f}}}J_{m+1}(2\pi\rho\rho_{\mathrm{f}})-\mathrm{e}^{-\mathrm{i}\phi_{\mathrm{f}}}J_{m-1}(2\pi\rho\rho_{\mathrm{f}})\right],
\end{align}

From these expressions, we can see that the focal field of a radially polarized phase vortex of topological charge $m$ has the following properties:
\begin{enumerate}
\item Its transverse focal field component is proportional to $(1-\rho^2s_0^2)^{1/4}$, while its longitudinal component is proportional to $s_0\rho/(1-\rho^2s_0^2)^{1/4}$. This means that for pupil fields with lower NA of which  lower spatial frequency part dominates, the transversally polarized component is stronger; while for a pupil function with higher NA of which the higher spatial frequency part dominates, the longitudinal component is stronger. This gives a way to modulate the polarization in the focal region.
\item The transverse focal field component of radially polarized $Z_n^{\pm1}$ is non-zero on the optical axis, due to the existence of $J_0$ in the integral.
\item The radial polynomial $R_n^{|m|}$ modifies the focal field distribution in such a way that for a fixed $m$ and $n>|m|$, two focal points appear symmetrically to the focal plane $z_{\mathrm{f}}=0$, and as $n$ increases, the separation between these two foci also increases, as shown in Fig. \ref{fig:radvaryingNM}a.
\item The width of the hollow spot of $I_z=|E_z|^2$ generated by radially polarized $Z_n^m$ increases with increasing $|m|\ne0$, as shown in Fig. \ref{fig:radvaryingNM}b.
\begin{figure}[htbp]
\centering
{\includegraphics[width=8.4cm]{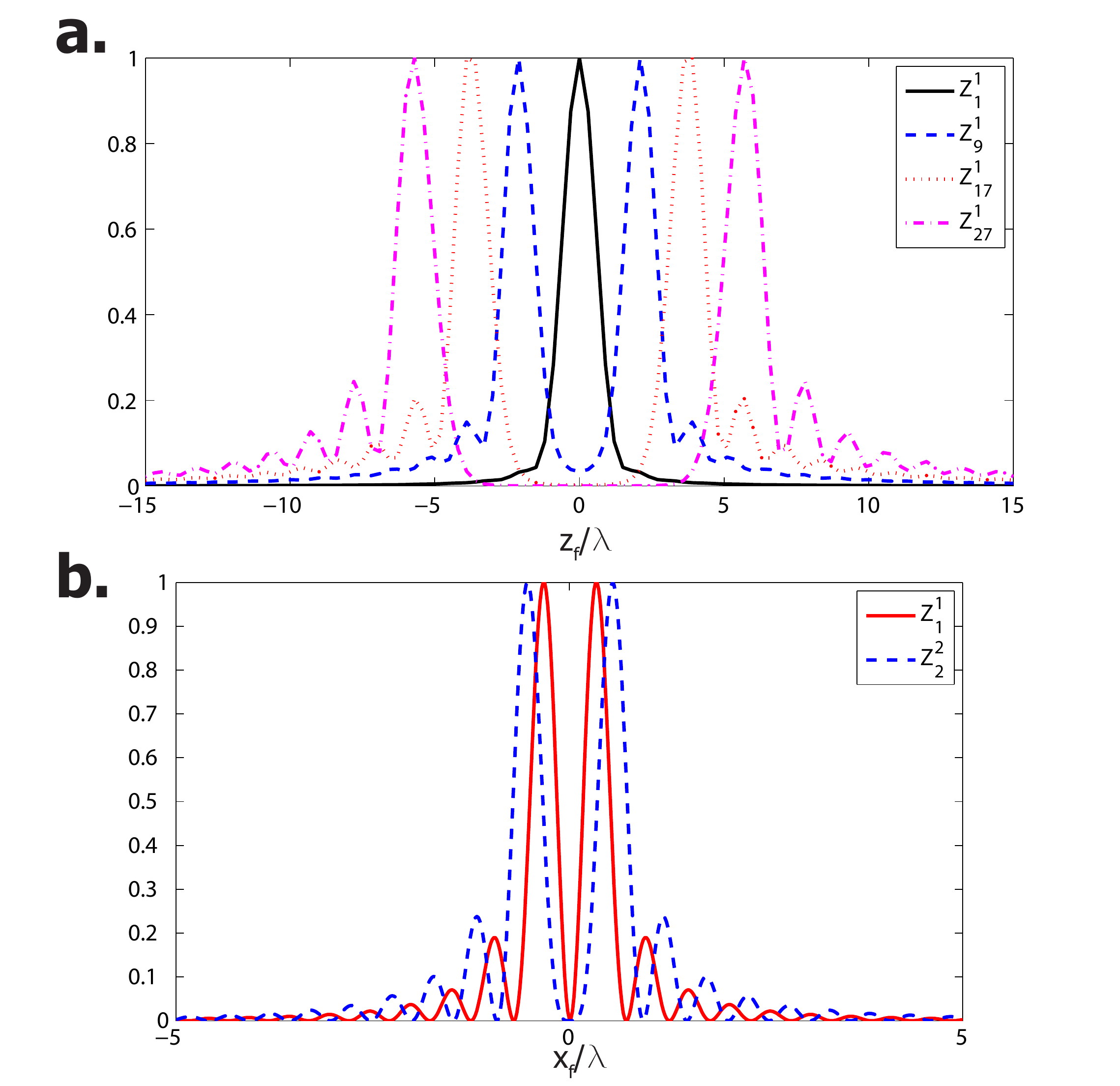}}
\caption{(a). Normalized z-component of the focal intensity for various radially polarized $Z_n^1$ along $z_{\mathrm{f}}$ axis, lateral position is at its intensity maxima in the transversal plane, the numerical aperture is 0.99. (b). Normalized z-component of the focal intensity for various radially polarized pupil field $Z_m^m$ along $x_{\mathrm{f}}$ axis, z position is at its intensity maxima in the longitudinal plane, the numerical aperture is 0.99. }
\label{fig:radvaryingNM}
\end{figure}
\end{enumerate}
While the focal field of an azimuthally polarized phase vortex of topological charge $m$ has the following properties:
\begin{enumerate}
\item The azimuthal polarization state is kept in the focal region for azimuthally polarized pupil field $Z_n^0$ , which can be easily derived from Eq. \ref{eq:azimex} and Eq. \ref{eq:azimey};
\item The transverse focal field component of $Z_n^{\pm1}$ is non-zero on the optical axis, due to the existence of $J_0$ in the integral, this can be confirmed by previous work by L.E. Helseth\cite{LEHelseth2004}.  And from Eq. \ref{eq:azimex} and Eq. \ref{eq:azimey}, the focal fields on the optical axis, i.e. $\rho_{\mathrm{f}}=0$, is circularly polarized.  For azimuthally polarized pupil field $Z_n^{\pm 1}$, the double focus phenomena along the optical axis also appears for $n>1$ and as $n$ increases, the separation between the foci increases.
\end{enumerate}

\section{Focal field shaping}
In this paper, we only use $Z_n^m$ with the same phase vortex charge $m$ for shaping the focus, because this makes the intensity circularly symmetric around the optical axis.  The basis of focal fields on a given grid in the $(x_\mathrm{f},z_\mathrm{f})$-plane corresponding to the radially and azimuthally polarized pupil fields $\vec{E}_{\mathrm{f}}$ are pre-calculated using the FFT and stored in the database.  The focal field of a general polarized electric pupil field can be written as as a linear combination of these basic functions:
\begin{equation}
\vec{E}_{\mathrm{f}}=\sum_{p}\beta_n^m \vec{E}^{nm}_{\mathrm{f}},
\end{equation}
where $m$ is fixed and $p=(n-|m|)/2$.\\

Firstly we use the transverse component of the focal intensity $I_t^{|m|m}(x_{\mathrm{f}},0)$ or longintudinal component of focal intensity $I_z^{|m|m}(x_{\mathrm{f}},0)$ to determine the lateral position $x_0$ where $I_t^{|m|m}(x_{\mathrm{f}},0)$ or $I_z^{|m|m}(x_{\mathrm{f}},0)$ is at its maximum. $\vec{E}^{nm}_{\mathrm{f}}(z_\mathrm{f})|_{x_{\mathrm{f}}=x_0}$ is then used to get a desired through-focus intensity distribution, where the Zernike coefficients $\beta_{n}^{m}$ are the optimisation variables.\\

Due to the fact that $\vec{E}^{nm}_{\mathrm{f}}$ has been precalculated, the focal field can be calculated rather fast. A random search of $\beta_n^m$ is applied in order to get a good starting point of the optimisation, then a Nelder-Mead simplex direct search method is used to optimize $\beta_n^m$ in order to get a minimum of $\left||\sum_p\beta_n^m\vec{E}^{nm}_{\mathrm{f}}|^2-I_{target}\right|$. Once the Zernike coefficients are found, the desired pupil function can be calculated easily by $\vec{E_s}=\hat{e_s}(\rho,\phi)\sum_{n,m}\beta_n^m Z_n^{m}$, where the polarisation of the pupil is either radially polarized that $\hat{e_s}(\rho,\phi)=\hat{e}_{\rho}(\rho,\phi)=(\cos\phi, \sin\phi)$, or azimuthally polarized that $\hat{e_s}(\rho,\phi)=\hat{e}_{\phi}(\rho,\phi)=(-\sin\phi, \cos\phi)$.
A few examples are given in the following section:

\subsection{Longitudinally polarized elongated subwavelength hollow channel}
An elongated subwavelength hollow channel can be used to trap absorbing nano-particles\cite{CDenz2012} and cold atoms\cite{MSZhan2010}. It can also be used as an excitation beam for high resolution fluorescence microscope like STED\cite{SWHell2006}, to excite fluorescence emission over an extended depth of focus, while keeping a high resolution.\\  

The z-component $E_{\mathrm{f},z}^{nm}$ of the focused field due to a radially polarized $Z^m_n$ pupil field for given fixed $m \neq 0$, vanishes on the optical axis, while  the topological charge of the incident beam $m$ is kept in $E_{\mathrm{f},z}^{nm}$.
 However its transverse components are non-zero on the optical axis. In the optimisation process, we therefore try to maximize the z-component by minimizing  $|I_z/I_{max}-I_{target}|$, where $I_z=|\sum_p\beta_n^1 E^{n1}_{\mathrm{f}z}|^2$, $I_{max}$ is the maximum intensity in the xz plane, and the target function $I_{target}$ is a rectangle function along $z_\mathrm{f}$ axis with value 1 between $|z_\mathrm{f}|<z_{\mathrm{max}}$, where $2\times z_{\mathrm{max}}$ is the desired depth of focus. \\

\begin{figure}[htbp]
\centering
{\includegraphics[width=8.4cm]{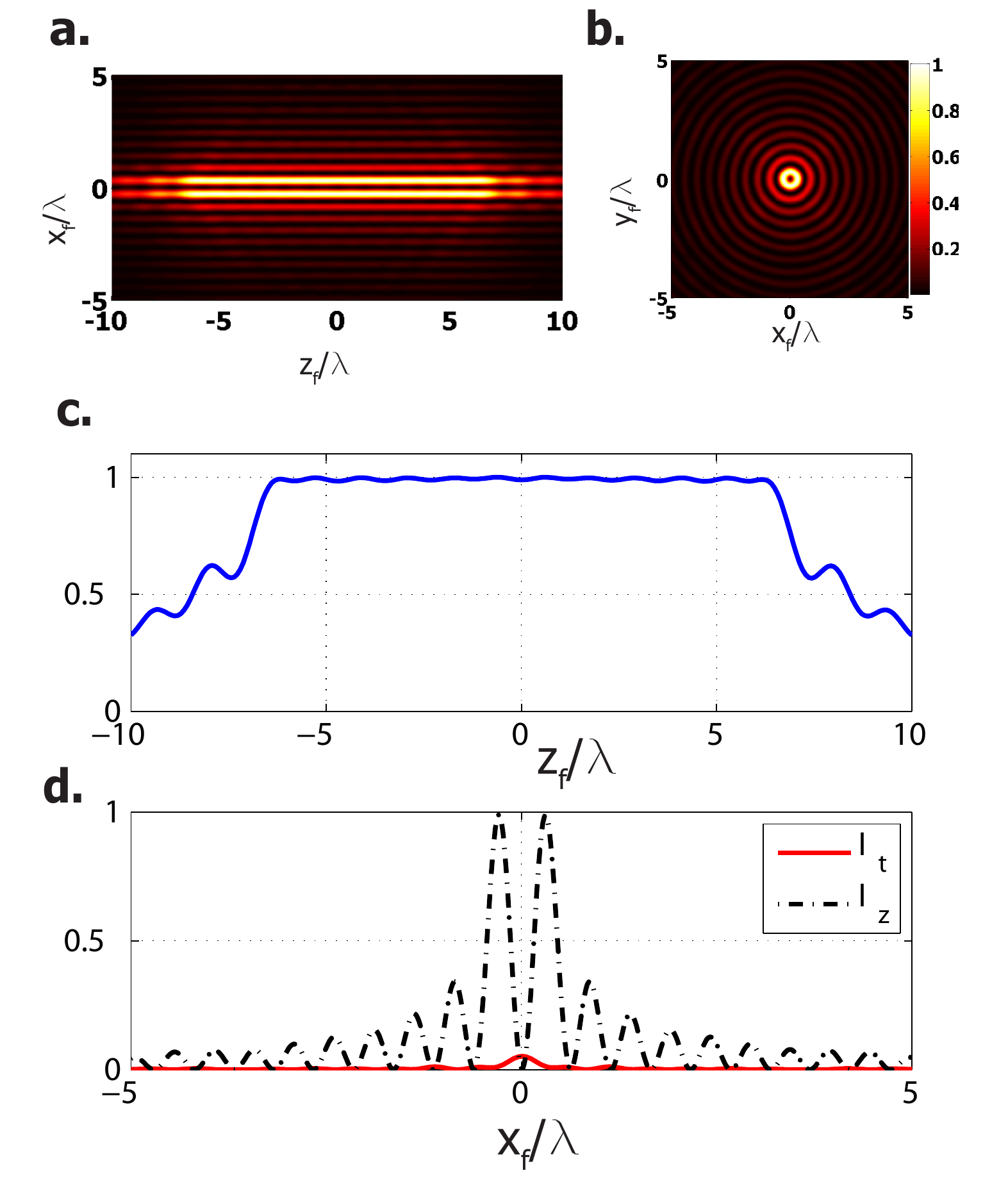}}
\caption{Elongated subwavelength hollow spot by an shaped radially polarized phase vortex with $m=1$ on the exit pupil, NA=0.99. (a).Nomalized focal field distribution in the $x_{\mathrm{f}}-z_{\mathrm{f}}$ plane; (b)Focal field in the transversal plane; (c). Focal intensity distribution along the $z_{\mathrm{f}}$ axis;, while $x_{\mathrm{f}}$ is at the maxmimum intensity on the transversal plane (d). Focal intensity distribution along $x_{\mathrm{f}}$ axis. }
\label{fig:radm1focal}
\end{figure}

\begin{figure}[htbp]
\centering
{\includegraphics[width=8.4cm]{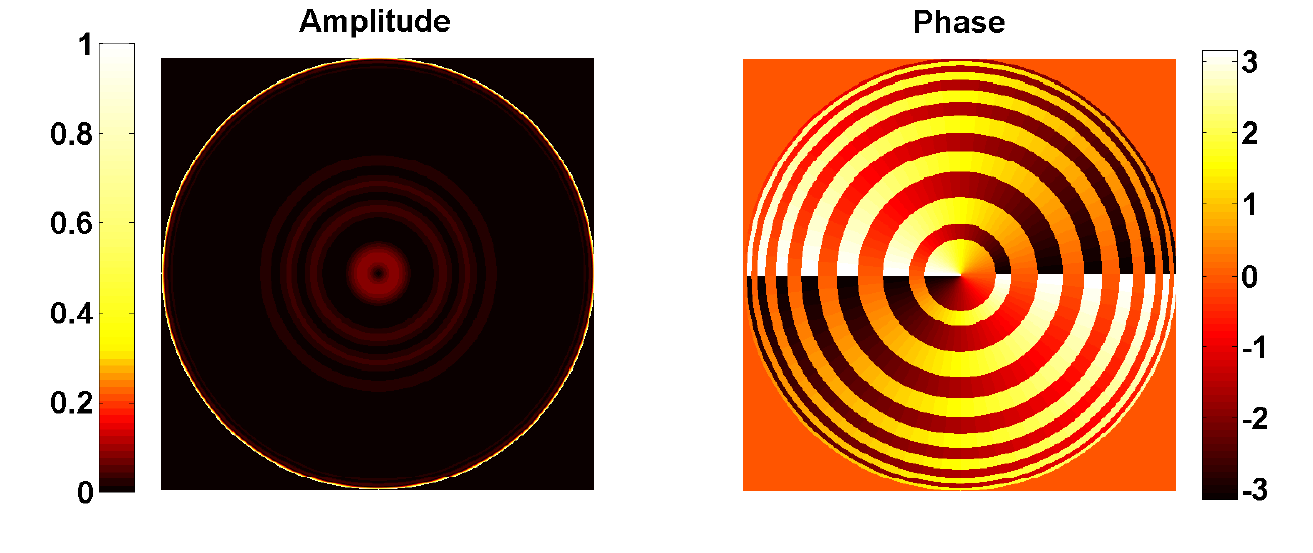}}
\caption{The pupil field of an shaped radially polarized phase vortex with $m=1$ to generate elongated subwavelength hollow spot in Fig. \ref{fig:radm1focal}.}
\label{fig:radm1pupil}
\end{figure}

We use a combination of 14 radially polarized Zernike polynomials $\hat{e}_{\rho}[Z_n^1]|_{n=1,3,5,...,27}$ on the exit pupil of an optical system with NA=0.99. As shown in Fig.\ref{fig:radm1focal}, with a set of Zernike coefficients $[\beta_n^1]|_{n=1,3,5,...,27}$= [-0.1437, -0.3, -0.398, -0.589, -0.568, -0.824, -0.674, -0.892, -0.785, -0.743, -0.754, -0.73, -0.182, -0.829], we get a hollow spot with lateral resolution(FWHM) of $0.28\lambda$, and its FWHM of focal depth is about $16\lambda$, with a good uniformity over a range of $12\lambda$ around the focal plane. For a high NA system with NA=0.99, this is equivalent to 12.24 Rayleigh unit $\lambda/NA^2$. This focal field is longitudinally polarized along the optical axis.  The corresponding optimized radially polarized pupil field with m=1  to generate this elongated hollow spot is given by $ \vec{E_s}(\rho,\phi)=\sum_{n}\hat{e_{\rho}}\beta_n^1 Z_n^{1}(\rho,\phi)$, as shown in Fig.\ref{fig:radm1pupil}.\\

\begin{figure}[htbp]
\centering
{\includegraphics[width=7.5cm]{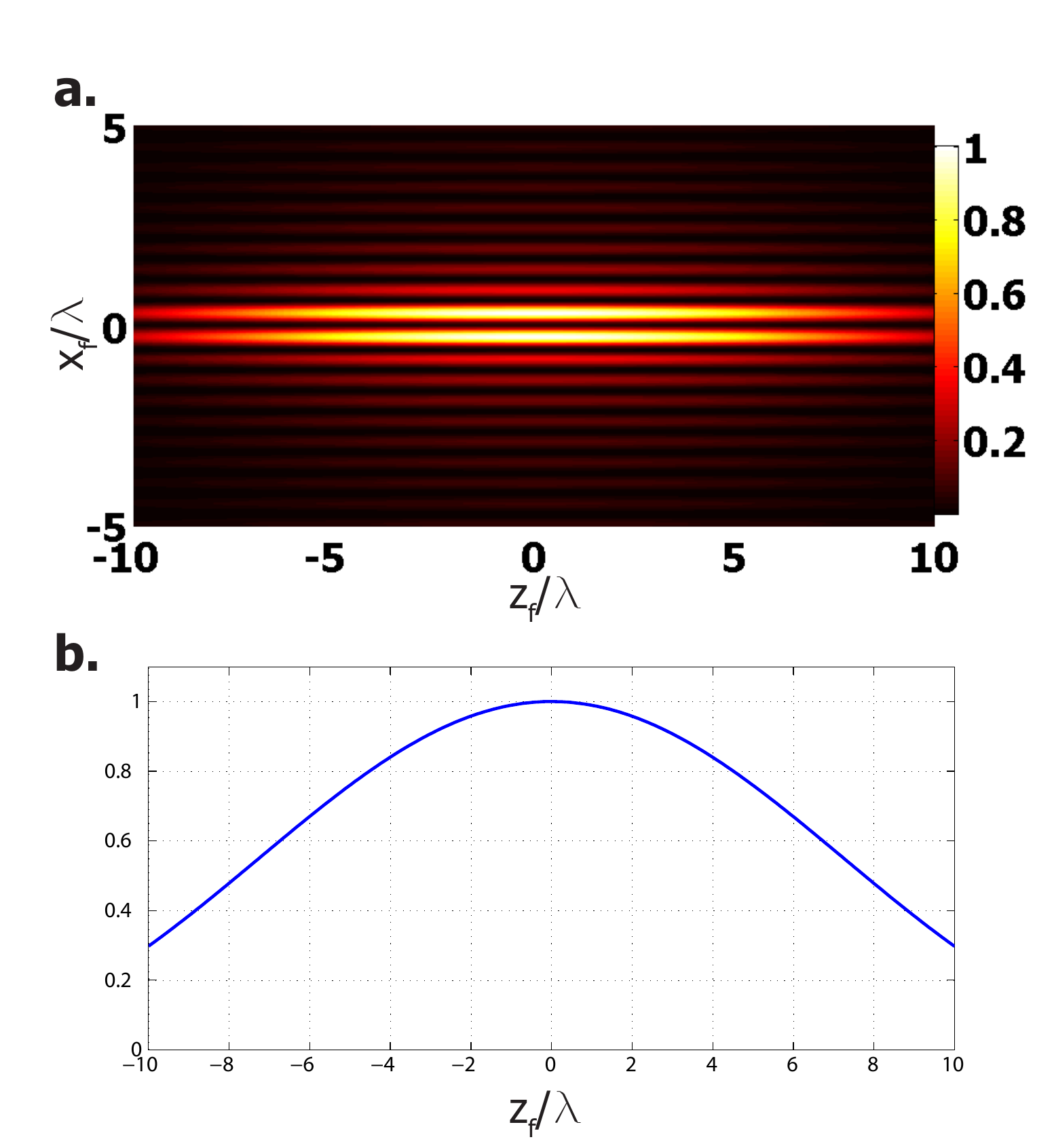}}
\caption{Elongated subwavelength hollow spot by a radially polarized annular aperture with phase vortex $m=1$ on the exit pupil, NA=0.99. (a).Nomalized focal field distribution in the $x_{\mathrm{f}}-z_{\mathrm{f}}$ plane; (b). Focal intensity distribution along the $z_{\mathrm{f}}$ axis;, while $x_{\mathrm{f}}$ is at the maxmimum intensity on the transversal plane }
\label{fig:annuradm1focal}
\end{figure}

The ratio of the transverse and longitudinal components can be adjusted by pupil engineering of the low spatial frequency and high spatial frequency components.  One commonly used method is to use annular aperture. For instance, a radially polarized annular pupil with phase vortex $m=1$: $A(0.99<\rho<1)=\hat{e}_{\rho}\exp(\mathrm{i}\phi)$  and 0 elsewhere in the exit pupil can also generate an elongated subwavelength hollow channel, as shown in Fig.\ref{fig:annuradm1focal}, but the uniformity of the spot along z dimension is much worse than the results obtained with our method.\\

\subsection{On-axis transversl polarized focal field with  radially polarized pupil $Z_n^1$}
The ability of modulating the local polarization of the focal field can be used to determine the orientation of a single optical emitter\cite{SWHell2001}.This can be achieved by engineering the pupil field. \\

 In the first example, we achieved an elongated hollow spot with radially polarized pupil field $Z_n^1$, for which at the focal points where the intensity is maximum, the  focal electric field is polarized primarily along the z-direction.  With the same set of Zernike polynomials, we can also maximize the tranverse compoents by optimizing the $\beta_n^1$ to minimize  $|I_t/I_{max}-I_{target}|$, where $I_t=|\sum_p\beta_n^1E^{n1}_{\mathrm{f}x}|^2+|\sum_p\beta_n^1E^{n1}_{\mathrm{f}y}|^2$, $I_{max}$ is the maximum total intensity in the xz plane,  the target function $I_{target}$ is a rectangle function along $z_\mathrm{f}$ axis with value 1 between $|z_\mathrm{f}|<z_{\mathrm{max}}$, where $2\times z_{\mathrm{max}}$ is the desired depth of focus..\\

\begin{figure}[htbp]
\centering
{\includegraphics[width=8.4cm]{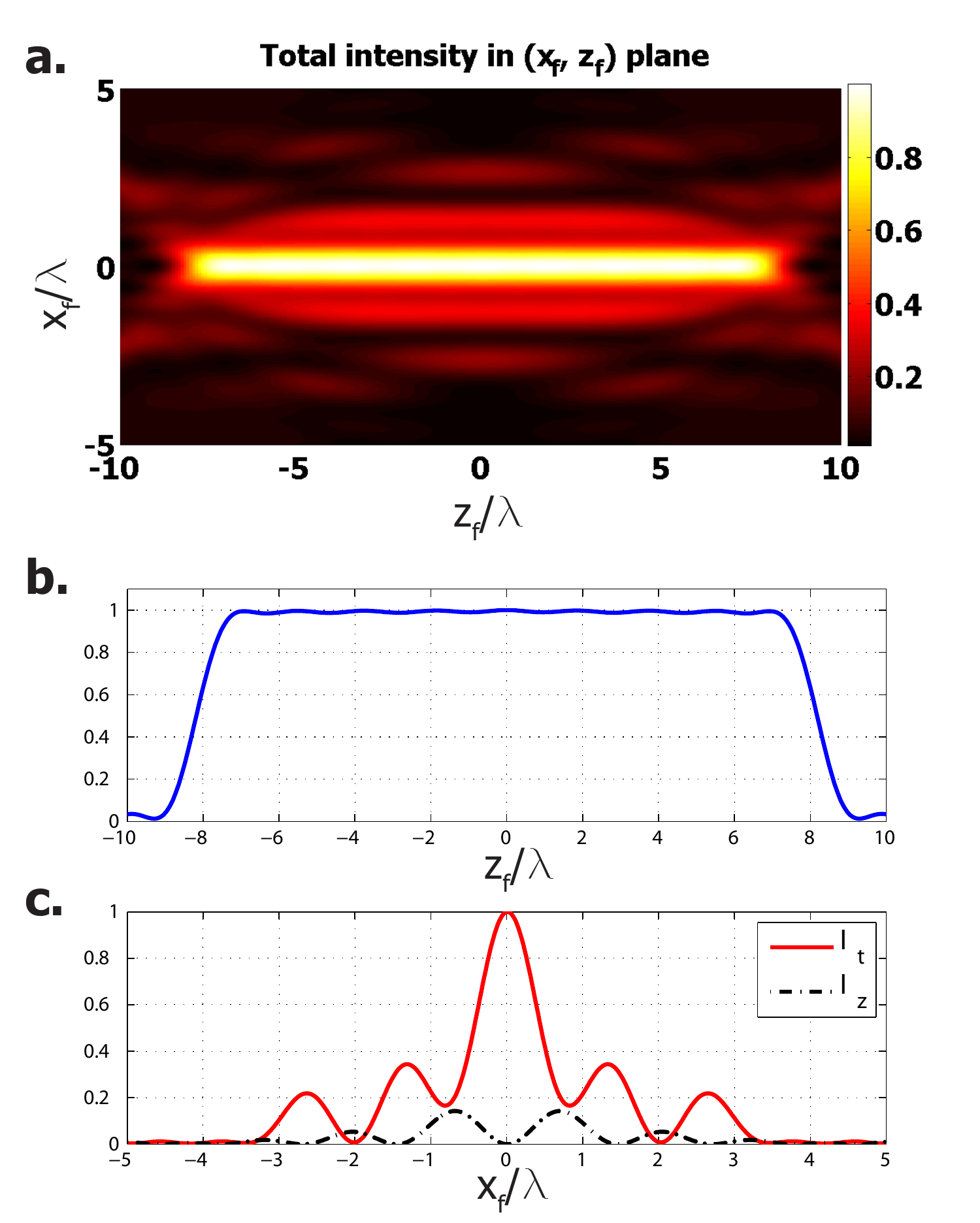}}
\caption{Elongated focal spot with stronger transversal polarisation component by an shaped radially polarized phase vortex with $m=1$ on the exit pupil, NA=0.9. (a).Nomalized focal intensity distribution in the $x_{\mathrm{f}}-z_{\mathrm{f}}$ plane;  (b). Focal intensity distribution along the optical axis; (c). Transversal and longintudinal components of the focal intensity distribution along $x_{\mathrm{f}}$ axis. }
\label{fig:radm1tfocal}
\end{figure}

We use a combination of 14 radially polarized Zernike polynomials $\hat{e}_{\rho}[Z_n^1]|_{n=1,3,5,...,27}$ on the exit pupil of an optical system with NA=0.9. As shown in Fig.\ref{fig:radm1tfocal}, with the set of Zernike coefficients $[\beta_n^1]|_{n=1,3,5,...,27}$=[-0.2632, 0.7, -0.677, -0.09, 0.982, -1.273, 0.413, 0.553, -1.167, 0.208, -0.194, 0.135, -2.244, 0.586], we get an elongated focal spot along the optical axis with strong polarization in the transverse direction, while its z-polarized component is much weaker. \\

\begin{figure}[htbp]
\centering
{\includegraphics[width=8.4cm]{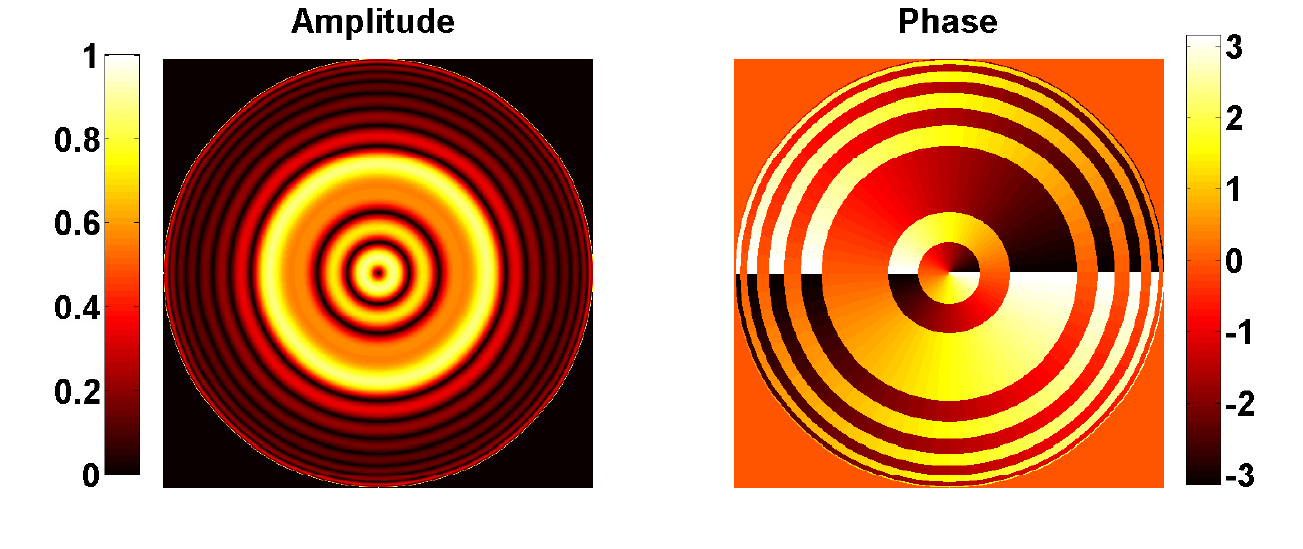}}
\caption{The pupil field of an shaped radially polarized phase vortex with $m=1$ to generate elongated transversally polarized focal spot in Fig. \ref{fig:radm1tfocal}.}
\label{fig:radm1tpupil}
\end{figure}

 The pupil field of an shaped radially polarized phase vortex with $m=1$ to generate this elongated tranversely polarized focal spot is then obtained from $ \vec{E_s}(\rho,\phi)=\hat{e_{\rho}}\sum_{n}\beta_n^1 Z_n^{1}(\rho,\phi)$, as shown in Fig.\ref{fig:radm1tpupil}.

\subsection{Multiple circularly polarized focal spots along the optical axis with azimuthally polarized $Z_n^1$}
Multiple subwavelength excitation focal spot along the optical axis can be very useful for fluorescence microscopy, as it can simultaneously excite fluorescence at multiple planes in the specimen. P.P. Mondal proposed a method using interference of two counter-propagating long depth of focus PSFs to generate multiple excitation spot PSF\cite{PPMondal2011}. However, this approach utilizes a 4pi configuration, which requires more complex optical setup than just shaping the pupil field. From Section 2, we know that for Zernike polynomials with a fixed $m$ and $n>|m|$, dual focus appear symmetrically to the focal plane $z_{\mathrm{f}}=0$, and as $n$ increases, the separation between these two focus increases. By assigning certain Zernike coefficients to Zernike polynomials $Z_n^m$ with different $n$, we can generate multiple axial focal spots by shaping the exit pupil.\\

\begin{figure}[htbp]
\centering
{\includegraphics[width=7.5cm]{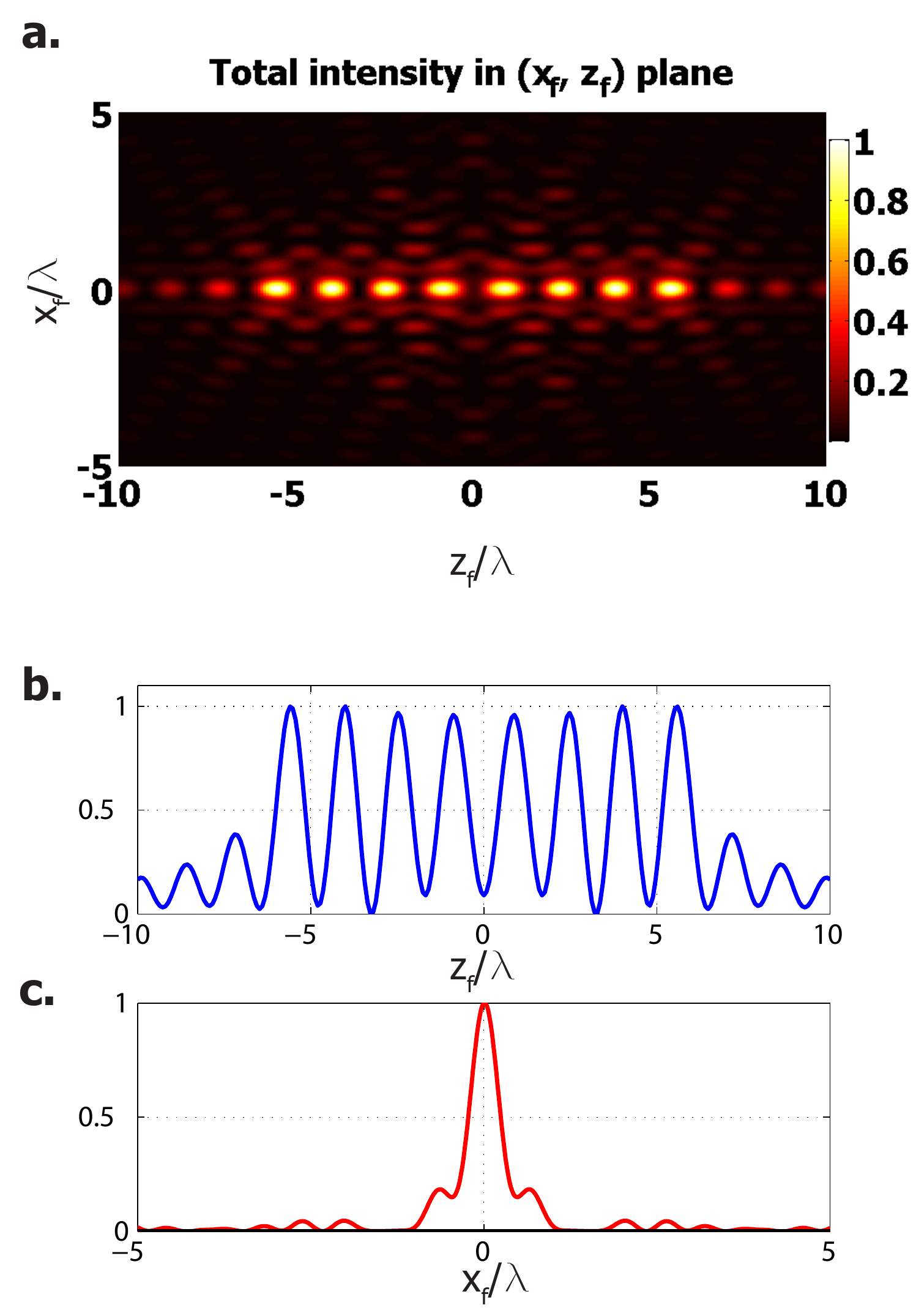}}
\caption{Multiple focal spot along optical axis by an shaped azimuthally polarized phase vortex with $m=1$ on the exit pupil, NA=0.99. (a).Nomalized focal field distribution in the $(x_{\mathrm{f}}, z_{\mathrm{f}})$ plane; (b). Focal intensity distribution along the optical axis; (c). Focal intensity distribution along the $x_{\mathrm{f}}$ axis, $z_{\mathrm{f}}$ is at one of the locations where the spot has maximum on axis intensity. }
\label{fig:azmum1focal}
\end{figure}

We use a combination of 14 azimuthally polarized Zernike polynomials $\hat{e}_{\phi}[Z_n^1]|_{n=1,3,5,...,27}$ on the exit pupil of an optical system with NA=0.99. As shown in Fig.\ref{fig:azmum1focal}, with the set of Zernike coefficients $[\beta_n^1]|_{n=1,3,5,...,27}$= [-0.222, 0.228, 0.4726, -0.106, 0.4315, 0.853, -0.169, -0.0438, 0.9424, 0.482, -0.85, 0.824, 0.557, 0.115], we can get 8 axial focal spots, each with a lateral resolution of $0.216\lambda$ and separated about $1.7\lambda$ between adjacent spots along the optical axis; The maximum intensity of these focal spots of the azimuthally polarized phase vortex of $m=1$ lies on the optical axis and contains no longitudinal component.\\

\begin{figure}[htbp]
\centering
{\includegraphics[width=8.4cm]{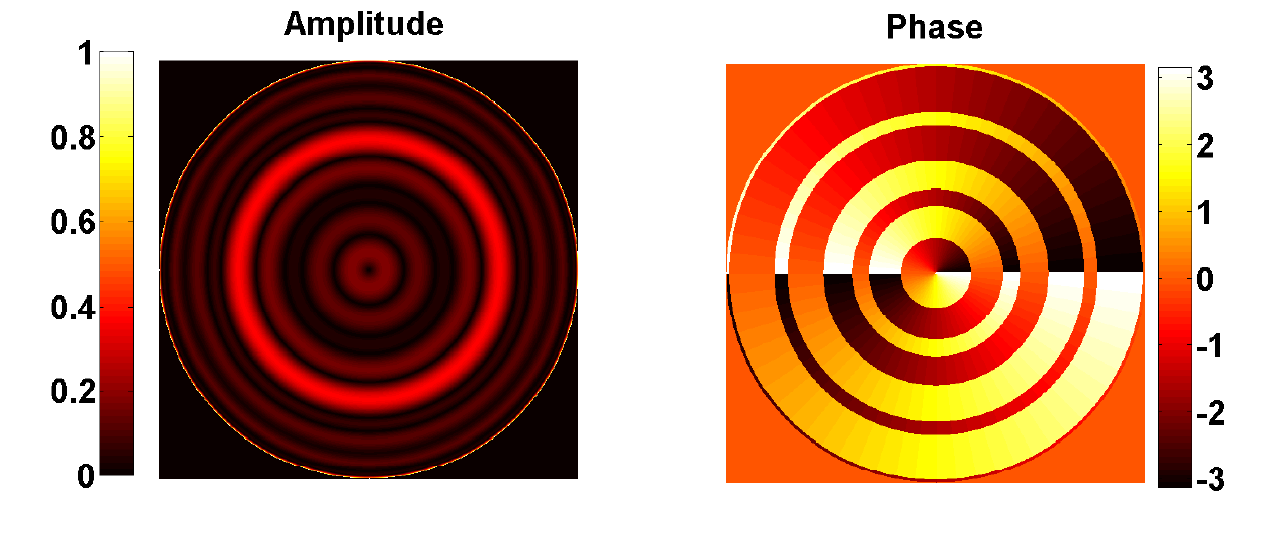}}
\caption{The pupil field of an shaped azimuthally polarized phase vortex with $m=1$ to generate multiple focal spot along optical axis.}
\label{fig:azmum1pupil}
\end{figure}

The pupil field of an shaped azimuthally polarized phase vortex with $m=1$ to generate this multiple focal spots along optical axis is obtained from $ \vec{E_s}(\rho,\phi)=\sum_{n}\hat{e}_{\phi}\beta_n^1 Z_n^{1}(\rho,\phi)$, as shown in Fig.\ref{fig:azmum1pupil}.

\section{Conclusion}
In this paper,the focal field of radially/azimuthally polarized complex Zernike polynomials is investigated.  Based on the understanding of this, a method to shape the focal field of radially/azimuthally polarized phase vortices is proposed. With pre-calculation of the focal field of each radially/azimuthally polarized Zernike polynomial, the optimization variables are reduced to the number of Zernike coefficients being used.  Three results from this method are given in this paper, including a longitudinally polarized subwavelength hollow focal spot  with a depth of focus up to $12\lambda$ and a lateral resolution of $0.14\lambda$ for a sytem with NA=0.99.  By engineering the ratio of high and low spatial frequencies, a tranversally polarization dominated elongated focal spot is obtained for radially polarized Zernike polynomials with topological charge of 1.  We also obtained  multiple subwavelength focal spots along the optical axis.  With these results,  pupil shaping of azimuthally/radially polarized optical vortices can achieve resolution improvement of optical system and can shape the polarization and intensity sistribution, in the focal region, which may result in interesting applications in the area of high resolution fluorescence microscopy, optical trapping, etc. 

% Bibliography


\newpage
\begin{thebibliography}{11}
\newcommand{\enquote}[1]{``#1''}

\bibitem{TGBrown00}
{ K. Youngworth and T.G. Brown}, \enquote{Focusing of high numerical aperture
  cylindrical vector beams,} Opt. Express \textbf{10}, 77--87 (2000).

\bibitem{QWZhan02}
{ Q.W. Zhan and J.R. Leger}, \enquote{Focus shaping using cylindrical vector
  beams,} Opt. Express \textbf{10}, 324--331 (2002).

\bibitem{QWZhan09}
{ Q.W. Zhan}, \enquote{Cylindrical vector beams: from mathematical concepts to
  applications,} Advances in Optics and Photonics \textbf{1}, 1--57 (2009).

\bibitem{WHan13}
{W. Han, Y.F. Yang, W. Cheng and Q.W. Zhan}, \enquote{Vectorial optical field
  generator for the creation of arbitrarily complex fields,} Opt. Express
  \textbf{21}, 20692--20706 (2013).

\bibitem{JPDing15}
{Z.Z. Chen, T.T. Zeng, B.J. Qian and J.P. Ding}, \enquote{Complete shaping of
  optical vector beams,} Opt. Express \textbf{23}, 17701--17710 (2015).

\bibitem{CSGuo14}
{Z.Y. Rong, Y.J. Han, S.Z. Wang and C.S Guo}, \enquote{Generation of arbitrary
  vector beams with cascaded liquid crystal spatial light modulators,} Opt.
  Express \textbf{22}, 1636--1644 (2014).

\bibitem{TVSPillai2012}
{K. Lalithambigai, P. Suresh, V. Ravi, K. Prabakaran, Z. Jaroszewicz, K.B.
  Rajesh, P.M. Anbarasan and T.V.S. Pillai }, \enquote{Generation of sub
  wavelength super-long dark channel using high na lens axicon,} Opt. Lett.
  \textbf{37}, 999--1001 (2012).

\bibitem{CDenz2012}
{C. Alpmann, M. Esseling, P. Rose and C. Denz }, \enquote{Holographic optical
  bottle beam,} App. Phy. Lett. \textbf{100}, 111101 (2012).

\bibitem{MSZhan2010}
{P. Xu, X.D. He, J. Wang and M.S. Zhan }, \enquote{Trapping a single atom in a
  blue detuned optical bottle beam trap,} Opt. Lett. \textbf{35}, 2164--2166
  (2010).

\bibitem{SWHell2006}
{K.I. Willig, S.O. Rizzoli, V. Westphal, R. Jahn and S.W. Hell }, \enquote{Sted
  microscopy reveals that synaptotagmin remains clustered after synaptic
  vesicle exocytosis,} Nature \textbf{440}, 935--939 (2006).

\bibitem{SWHell2001}
{N. Huse, A. Schonle and S.W. Hell }, \enquote{Z-polarized confocal
  microscopy,} J. Biomed. Opt. \textbf{6}, 480--484 (2001).

\bibitem{LNovotny2015}
{M. Kasperczyk, S. Person, D. Ananias, L.D. Carlos and L. Novotny },
  \enquote{Excitation of magnetic dipole transitions at optical frequencies,}
  Phys. Rev. Lett. \textbf{114}, 163903 (2015).

\bibitem{HFWang08}
{H.F. Wang, L.P. Shi, B. Lukyanchuk, C. Sheppard and C.T. Chong },
  \enquote{Creation of a needle of longitudinally polarized light in vacuum
  using binary optics,} Nature photonics \textbf{2}, 501--505 (2008).

\bibitem{JMWang2010}
{J.M. Wang, W.B. Chen and Q.W. Zhan}, \enquote{Engineering of high purity
  ultra-long optical needle field through reversing the electric dipole array
  radiation,} Opt. Express \textbf{18}, 21965--21972 (2010).

\bibitem{JMWang2013}
{J.M. Wang, Q.L. Liu, C.J. He and Y.W. Liu }, \enquote{Reversal construction of
  polarization-controlled focusing field with multiple focal spots,} Optical
  Engineering \textbf{52}, 048002 (2013).

\bibitem{YZYu2015}
{Y.Z. Yu and Q.W. Zhan }, \enquote{Creation of identical multiple focal spots
  with prescribed axial distribution,} Sci. Reports \textbf{5}, 14673 (2015).

\bibitem{LEHelseth2004}
{L.E. Helseth }, \enquote{Optical vortices in focal regions,} Opt. Commun.
  \textbf{229}, 85--91 (2004).

\bibitem{CJRSheppard2014}
{C.J.R. Sheppard }, \enquote{Polarized focused vortex beams: half-order phase
  vortices,} Opt. Express \textbf{22}, 18128--18141 (2014).

\bibitem{XLiu2010}
{X. Hao, C.F. Kuang, T.T. Wang and X. Liu}, \enquote{Phase encoding for sharper
  focus of the azimuthally polarized beam,} Opt. Lett. \textbf{35}, 3928
  (2010).

\bibitem{YKozawa2014}
{Y. Kozawa and S. Sato }, \enquote{Dark-spot formation by vector beams,} Opt.
  Letters \textbf{33}, 2326--2328 (2008).

\bibitem{YQZhao07}
{Y.Q. Zhao, J.S. Edgar, G.D.M. Jeffries, D. McGloin and D.T. Chiu },
  \enquote{Spin-to-orbital angular momentum conversion in a strongly focused
  optical beam,} Phy. Rev. Letters \textbf{99}, 073901 (2007).

\bibitem{LWei2014}
{AP Konijnenberg, L Wei, N Kumar, L Cisotto, SF Pereira, HP Urbach},
  \enquote{Demonstration of an optimised focal field with long focal depth and
  high transmission obtained with the extended nijboer-zernike theory,} Opt.
  Express \textbf{22}, 311--324 (2014).

\bibitem{JJMBraat08}
{J.J.M. Braat, S. van Haver, A.J.E.M. Janssen and P. Dirksen},
  \enquote{Assessment of optical systems by means of point-spread functions,}
  Progress in Optics \textbf{51}, 349--468 (2008).

\bibitem{SHaver2013}
{S. van Haver and A.J.E.M. Jassen}, \enquote{Advanced analytical treatment and
  efficient computation of the diffraction integrals in the extended nijboer
  zernike theory,} J. Europ. Opt. Soc. Rap. Public. \textbf{8}, 13044 (2013).

\bibitem{EWolf1958}
{B. Richards and E. Wolf}, \enquote{Electromagnetic diffraction in optical
  systems ii. structure of the image field in an aplanatic system,} Proc. R.
  Soc. Lond. A \textbf{253}, 358--379 (1959).

\bibitem{TLasser2006}
{M. Leutenegger, R. Rao, R.A. Leitgeb and T. Lasser}, \enquote{Fast focus field
  calculations,} Opt. Express \textbf{14}, 11277--11291 (2006).

\bibitem{PPMondal2011}
{P.P. Mondal and A. Diaspro}, \enquote{Simultaneous multilayer scanning and
  detection for multiphoton fluorescence microscopy,} Sci. Reports \textbf{1},
  149 (2011).

\end{thebibliography}
\end{document}